# Multi-Method Li Plating Characterization of a Commercial 26 Ah Li-Ion Pouch-Cell

Christiane Rahe,[1,2,3]* Heinrich Ditler,[1,2,3] Thorsten Tegetmeyer-Kleine,[1,2,3] Marius Flüge[1,4] Thomas Waldmann,[4,5] Margret Wohlfahrt Mehrens,[4,5] Philipp Schleker,[6] Peter Jakes,[6] Beatrice Wolff,[6] Josef Granwehr,[6,7] Rüdiger-A. Eichel,[3,6,8] Jiří Vacík,[9] Giovanni Ceccio,[9] Antonino Cannavo,[10] Ivana Pivarníková,[11,12] Ralph Gilles,[11] Peter Müller-Buschbaum,[12] Adrian Mikitisin,[13] Joachim Mayer,[13] Michael Noyong,[14] Ulrich Simon,[14] Marius Bolsinger,[15] Volker Knoblauch,[15] and Dirk Uwe Sauer[1,2,3,16]

[1]*Institute for Power Electronics and Electrical Drives (ISEA), RWTH Aachen University, 52074 Aachen, Germany*
[2]*Center for Ageing, Reliability and Lifetime Prediction of Electrochemical and Power Electronics Systems (CARL), RWTH Aachen University, 52074 Aachen, Germany*
[3]*Jülich Aachen Research Alliance - JARA Energy, RWTH Aachen University, Templergraben 55, 52056 Aachen, Germany*
[4]*Zentrum für Sonnenenergie- und Wasserstoff-Forschung Baden-Württemberg (ZSW), Helmholtzstraße 8, 89081 Ulm, Germany*
[5]*Helmholtz Institute Ulm for Electrochemical Energy Storage (HIU), 89081 Ulm, Germany*
[6]*IET-1, Forschungszentrum Jülich, Wilhelm-Johnen-Straße, 52428 Jülich, Germany*
[7]*Institute of Technical and Macromolecular Chemistry (ITMC), RWTH Aachen University, 52074 Aachen, Germany*
[8]*Institute of Physical Chemistry and Faculty of Mechanical Engineering, RWTH Aachen University, 52074 Aachen, Germany*
[9]*Nuclear Physics Institute Czech Academy of Sciences (NPI), CZ-25068, Řež, Czech Republic*
[10]*Brookhaven National Laboratory (BNL) Collider Accelerator Department, Upton, NY 11973-5000, United States of America*
[11]*Heinz Maier-Leibnitz Zentrum (MLZ), Technische Universität München, Lichtenbergstr. 1, 85748 Garching, Germany*
[12]*Technical University of Munich, TUM School of Natural Sciences, Department of Physics, Chair for Functional Materials, James-Franck-Str. 1, 85748, Garching, Germany*
[13]*Gemeinschaftslabor für Elektronenmikroskopie (GfE), RWTH Aachen, Ahornstraße 55, 52074 Aachen, Germany*
[14]*Institut für Anorganische Chemie (IAC), RWTH Aachen, Landoltweg 1A, 52074 Aachen, Germany*
[15]*Aalen University of Applied Sciences (AU), Materials Research Institute, 73430 Aalen, Germany*
[16]*Helmholtz Institute Münster (HIMS), Forschungszentrum Jülich, Wilhelm-Johnen Straße, 52428 Jülich, Germany*
* correspondig author

## Abstract

Lithium (Li) plating on graphite is a significant degradation mechanism in Li-ion batteries. While numerous experimental techniques have been used to study Li plating in laboratory cells, investigations of commercial high-energy cells often rely on electrochemical methods. Here we present and classify various methods for detecting Li plating on a commercial A123 pouch cell. In a round robin study across multiple battery research laboratories, Li-plated graphitic electrode material was analyzed using electrochemical, microscopic, and spectroscopic methods capable of detecting metallic Li deposits. After cell opening, their overall distribution on the anode surface was examined using a flatbed scanner to ensure comparability of the samples. Optical and electron microscopy provided detailed surface and, in combination with a focused ion beam, subsurface structure and morphology. Spectroscopic methods confirmed the presence and onset of plated Li with varying sensitivity. Moreover, spectroscopic and imaging techniques were combined correlatively where possible. Availability and measurement duration of each technique was compared. Optical methods are fast and easy to use; thus, they are recommended for most samples, with spectroscopic confirmation reserved for reference samples. This multimodal study demonstrates a range of methods that can be used alone or in combination to qualitatively or quantitatively detect Li-plating.

## Introduction

Today, Li-ion batteries are widely used as energy storage devices. The energy transition is accelerating their use to meet the goal of increasing energy production from renewable sources, reducing exhaust emissions, and the associated reduction in the use of fossil fuels.

However, Li-ion batteries still present challenges in terms of lifetime and safety. Li plating is one of the main aging factors, and at the same time posing a safety risk.[1,2]

Li plating occurs when Li-ion intercalation into the negative electrode material is energetically or kinetically disadvantageous due to environmental conditions.[1,2] This results in the formation of metallic Li, which is either passivated, reintercalated or stripped again in the next discharge step. The passivation of metallic Li leads to an aging effect of the batteries due to the loss of cyclable Li, which is no longer available to store energy in the next cycle, and thus a loss of capacity.

Depending on the position and surface area in the cell, further capacity loss may be caused by the deactivation of active material. Therefore, from an application point of view, Li plating should be avoided.

Moreover, Li plating can also lead to safety-critical conditions.[1] Metallic dendrites can penetrate the separator and establish a conductive connection between the anode and cathode. Ideally, the dendrite breaks, or the separator closes in the heat generated by the high current flow. Additionally, exothermic reactions by plated Li with electrolyte[3–5] can lead to cell heating if sufficient heat dissipation is not possible.[5–8] In the worst case, both internal short circuits and exothermic reactions of Li metal can lead to a thermal runaway.

In the past, studies delivered evidence for Li-plating based only on few methods each, with many methods solely indicating the presence of metallic Li.[1] Most commonly, electrical data was used by analyzing the stripping discharge[9,10] or voltage relaxation curve.[11,12] Intercalation mechanisms were evaluated via changes of $LiC_x$ Bragg reflections studied from neutron diffraction.[13] Other research groups opened the cells and examined the deposits on the graphite anode surface under the microscope[14,15] (morphology and color) or SEM[7] (morphology).

Kayser et al.[16] showed on a lab cell, that nuclear magnetic resonance (NMR) provides quantitative relative information on electrochemically active and of "dead" Li in operando. Electron paramagnetic resonance (EPR) was used to detect Li-dendrites in battery separators.[17]

In this work, we present different methods for detecting Li plating. Early detection and a deep understanding of Li plating's occurrence are necessary to produce more reliable batteries in the future and/or to design application strategies accordingly.

By comparing a variety of methods from different laboratories, their suitability and limitations to detect Li plating are investigated. This knowledge can help in the future to select specific analysis methods and, at the same time, refer to the reliability of the methods through this publication.

The aim of this work is to present and compare methods for detecting metallic lithium in batteries so that further cells can be investigated on the basis of this work using methods that have already been verified.

In the summary of this paper, the methods are sorted according to their spatial resolution. One unique feature of this paper is that all results presented have been obtained on the same cell, thus ensuring high comparability of the measurement results. Individual methods have also been performed in different laboratories to verify the accuracy of the measurements.

**Experimental**

**The investigated cell** is an A123 26 Ah battery (WX1413726). The cell format is a pouch bag with 161 × 227 × 7.5 mm dimensions. The active material is NMC 111 for the cathode and

graphite for the anode. A short summary of the A123 26 Ah cell parameters is shown in **Table I.**

| Parameter | Value |
| --- | --- |
| Nominal Capacity | 26 Ah |
| Nominal Voltage | 3.7 V |
| Voltage Range | 2.8–4.15 V |
| Active Material | NMC 111 vs Graphite |
| Cell Dimensions | 161 × 227 × 7.5 mm |
| Cell Weight | 550 g |
| Specific Energy | 175 Wh kg$^{-1}$ |
| Anode Dimensions | 20 sheets à 148 × 198 mm |
| Anode Thickness (double-sided) | 173 μm |
| Cu Current Collector Thickness | 10 μm |
| Cathode Dimensions | 19 sheets à 145 × 195 mm |
| Cathode Thickness (double-sided) | 150 μm |
| Al Current Collector Thickness | 20 μm |
| Separator Thickness | 18–20 μm |

The cell selection process ensured that the cell met the following criteria: medium size, commercially produced, stacked pouch, and a graphite anode with no silicon content.

The medium cell size is used to examine the Li plating distribution on the electrode surface in more detail, and the stacked electrodes are ideal to be scanned with a flatbed scanner.

Furthermore, stack production is important to us in order to rule out further changes/effects caused by winding and different pressures in the stack. For this study, we are focusing on the possibilities of investigating Li plating on graphite. The silicon content in modern cells contributes to higher cell capacities, but due to the sequence of use in terms of charging/discharging, graphite is responsible for the occurrence of plating in relation to the distance from 0 V. However, silicon in the electrode leads to further changes, such as volume expansion, associated material loss, and recrystallisation of the silicon. These effects should not limit the search for metallic Li detection methods, which are the focus of this study. Cathode chemistry was not a focus of the cell selection process, as the anode is the focus of the investigations. A reliable, well-known cathode material was selected here, which promises stability and a sufficient number of Li ions.

**Li plating charge—electrical set-up**

The cell was electrically cycled using a Neware BTS-4016–5V100A battery cycler in a temperature-controlled climate chamber (Binder MK 53). To additionally monitor the temperature during all electrical tests, a type-K thermocouple was centrally placed on the upper cell surface. All electrical pre-tests and the following Li plating charge were conducted without a mechanical clamping construction and, therefore, without a passive cell cooling system.

Each cell was initially characterized at 25 °C with a C/10 cycle (constant current, constant voltage (CCCV) charge; constant current (CC) discharge) between 2.8–4.15 V to estimate the initial capacity. Unless stated otherwise, the CCCV phase was terminated at a C/50 current.

After the initial check-up, the cells were brought to their desired temperature, followed by a rest period of at least 5 h to ensure complete temperature homogenization.

To determine the Li plating behavior, a stripping differential voltage analysis (DVA) was conducted following the method described by Petzl et al.[18] For this purpose, a small C/20 discharge current was applied directly after (fast) charging the cells. The measured voltage during the slow discharge is derived with respect to the transferred charge (dV/dQ), and if reversible Li plating is present, a characteristic voltage plateau is visible.[9,10,18]

For the actual DVA test, two charge/discharge steps were conducted sequentially. The first step was to determine the cell capacity at the desired temperature, and the second step aimed to assess whether the cell plates when the total charge is loaded into it without a voltage limit.

The first cycle was conducted with a C/10 CCCV-charge to 4.15 V followed by a C/10 CC-discharge from 4.15 V to 2.8 V. The measured extracted capacity in Ah was used for the second cycle, which was carried out at the target C-rate, without voltage limitation but with capacity limitation.

The constant capacity charge ensured that the test was performed at a defined state of charge (SoC) for any temperature and C-rate. After the constant capacity charge, the previously described C/20 stripping discharge was conducted.

First tests were performed at C-rates ranging from C/10 to 1 C at 25 °C. Since no Li plating is expected at 25 °C and 1 C charge, these measurements are considered a reference for the following cold charge cycle, performed at −5 °C and 1 C.

The Li plating cycle (constant capacity charge) with which the cell is charged immediately before opening was 1 C at −5 °C. Special emphasis was placed on opening the cell as quickly as possible after fast charging to prevent changes in the expected coating, e.g., due to re-intercalation or reactions with the electrolyte.[11,13,19]

These voltage limits are far outside the specified range, and therefore, the cell should not be operated permanently in this voltage window since other aging effects, such as electrolyte degradation/gassing, can occur in addition to Li plating.[20]

**Cell opening**

All cell openings were performed immediately after the last cell charge (<10 min rest period) to ensure that the samples were mainly examined unchanged and with minimal time for re-intercalation of Li-ions. Opening and following sample preparation was done in an Ar filled glove box with low O2 (<5 ppm) and H20 (<0.3 ppm) values to ensure that the samples did not react with the environmental gas.

Immediately after opening and separating the cell components, the electrodes (anode/cathode) were analyzed with a flatbed scanner (Canon CanoScan LiDE 300) inside the glovebox. All electrodes were scanned individually from both sides using the flatbed scanner and the corresponding software (CanonScan LiDe 300). The scan resolution was set to the highest level, corresponding to 600 DPI and thus a pixel size of 42 μm.

After scanning the electrodes, they were individually packed in airtight aluminum bags to be transported to the other institutes for the following measurements. Care was taken to ensure that all partners received a sheet from the middle and top of the electrode stack.

To gain more insight into the Li plating behavior at the cell level, the scans of the flatbed scanner were analyzed using image processing techniques. This study involved the application of various computer vision (CV) and deep learning (DL) methods to analyze and extract relevant information.

The initial step for image processing is image acquisition. The raw images from the flatbed scanner served as the foundation for subsequent processing steps.

First, a dedicated Region of Interest (ROI) detection algorithm was deployed to identify and isolate individual sheets or specific regions of interest within the acquired images.

The detection and isolation of regions of interest (ROIs) in the acquired images was carried out using a dedicated algorithm for detecting rectangular objects, specifically rectangles. This algorithm combines image preprocessing techniques, such as median blurring, binary thresholding, and morphological operations, to extract contours of rectangle objects. After filtering, the detected contours are checked for size criteria, and the ROIs are cropped from the original image. These extracted ROIs are then further segmented using a supervised deep learning-based approach, employing a U-Net model.[21]

A deep learning-based segmentation approach was used to address overhang artifacts present in the ROI images. This involved manual labeling of a representative set of images, followed by the training of a +segmentation model. This model effectively removed the unwanted overhang from the ROI images, producing a cleaner representation of cell active material.

In the segmentation process, the nearest neighbor (NN) algorithm was used, which classifies values into k groups. The k-NN anchor values were manually established and utilized as thresholds, which were then used to generate masks for each individual sheet. This compression resulted in a Li plating intensity scale with four k*-values: 0 for overhang, 1 for lithiated graphite areas, 2 for light plating, and 3 for strong plating. This new representation enhanced the analysis of Li plating intensity across all electrode sheets.

To visually represent the distribution and concentration of Li plating in different regions across the sheets, each ROI was overlaid with its corresponding mask, creating a heatmap. In addition to sheet-level analysis, the Li plating distribution at the cell level can also be represented by a 3D representation, where colored masks of the individual sheets are superimposed.

**Microscopic methods**

Microscopic techniques use the properties of light to create images and to visualize and analyze objects with remarkable clarity. In this work, different microscopes (scanner, light microscope, laser scanning microscope, scanning electron microscope (SEM), X-ray microscope) were used for different magnifications and representations.

High-resolution **light microscopy** was performed with a ZEISS Axio Observer 7 with an EC Epiplan-Neofluar 100× objective using a neutral white (3200 K) LED and an extended depth of field mode inside the glovebox. The images were taken starting at the edge and moving toward the center of the electrode.

For the **laser scanning microscope** studies, a Keyence VK-X1100 was used. This microscopy system incorporates high-resolution imaging and analysis, including depth of field enhancement, color overlay imaging, and a 0.1 nm depth resolution. Samples of the electrode with an area of approx. 1 cm$^2$ was cut out from different regions and then fixed with double-sided carbon tape to an SEM holder. The images shown here were taken with 50× and 150× lenses. The microscope was located in the same Ar-filled glove box used to open the cells, which minimized transfer times and ensured that the specimens were not affected by moisture or oxygen. Using the SEM holder for the laser scanning microscope analysis allows the same sample to be subsequently examined in the SEM.

The **scanning electron microscope** (SEM) analysis was performed with a Zeiss Supra 55 using the secondary electron (SE)-detector and 5 kV accelerating voltage. The samples were transferred to the electron microscope using a vacuum transfer shuttle (Kammrath & Weiss GmbH) to protect the samples from oxygen and moisture at all times, mitigating the risk of unwanted reactions during the transfer. The SEM samples are the same as in the laser scanning microscope study.

A second SEM analysis was done with a JEOL JSM-IT800(HL) at 5 kV accelerating voltage and the SE detector from JEOL, as well as a backscattered electron (BSE) detector from Gatan.

In addition to the classic top/down SEM imaging, a **focused ion beam scanning electron microscope** (FIB-SEM) (Zeiss Crossbeam 350) was used to gain insight into the depth of the electrode material at an image resolution of an SEM (also using the SE-detector, 5 kV accelerating voltage). The cutting gallium ion beam removes material, so that deeper layers can be examined. The first rough cut was done with 15 nA and the polishing step with 3 nA. The imaging step was done with 30 kV accelerating voltage. In this study, a wedge was cut into the sample surface in two steps (rough cut and polishing step) to examine the surface's depth profile at the resulting cutting edge. Similar to the SEM investigation, a vacuum transfer shuttle (Semilab Inc.) was used from the glovebox to FIB-SEM to avoid contact with the atmosphere.

**X-ray microscopy** (XRM) was carried out using a ZEISS Xradia 620 Versa microscope. The X-ray microscope converts X-rays into visible light to enable further magnification with glass lenses for higher magnification compared to a CT. This approach achieves sub-micron resolution in micro-CT scanners. The samples were measured using Kapton tape as protective cover to protect the samples from the ambient air. For this high resolution the sample size was smaller than 5 mm in depth. An acceleration voltage of 60 kV (108 μA, filter: without, 12 s/image) and the 20× objective were selected for these recordings. The pixel size of the images is 0.5 um.

**Spectroscopy**

Spectroscopy methods are indispensable tools in the field of analytical chemistry, offering valuable insights into the composition, structure, and properties of matter. Compared to optical methods, spectroscopic methods are directly quantifiable and provide information about the material composition. Therefore, in the following different spectroscopy methods are presented, with which Li plating on a graphite anode can be examined.

The Zeiss Supra 55 SEM is additionally equipped with an **energy-dispersive X-ray** (EDX) spectroscopy detector - the windowless Oxford Instruments Ultim Extreme. This special EDX detector can directly detect the very weak Li signal (ultra-soft Li Kα X-rays at around 54 eV [21]) in the spectrum. Therefore, it is possible to validate the elemental compositions of the plated surface structures. Here, the accelerating voltage was 5 kV, which is relatively low for EDX measurements. However, preliminary investigations with different accelerating voltages showed that the best results for Li detection could be obtained with this accelerating voltage.

The Zeiss Crossbeam 350 SEM is equipped with an Oxford Instruments Ultim Max 100, which is also a state-of-the-art EDX detector with very high sensitivity, but cannot detect Li directly. Instead it was used to detect C, P, F, O. The accelerating voltage used was 10 kV.

The JEOL JSM-IT800(HL) is equipped with an Octane Elect EDX system from EDAX, a **soft X-ray Spectrometer** (SXES) SS-94000 from JEOL, and an InLux/Virsa **Raman Spectrometer** from Renishaw. These spectrometers allow the co-localized analysis inside the SEM. The EDX is limited to elements above Z = 4. The combination of the BSE from Gatan and EDX from EDAX enables the method of Österreicher et al.[22] to quantify Li in the SEM by the composition-by-difference (Li-CDM). In brief, first, a calibrated backscattered electron image (qBSE) was prepared by adjusting the dynamic range (contrast and brightness) of the detector to high and low Z elements (Z = 1 to Z = n). The BSE grayscale was calibrated by measuring the intensities of several standard materials with known Z numbers. Then, a quantitative EDS (standardless) mapping was recorded. During post-processing (using "Cipher" in the Digital Micrograph Suite, Gatan), the quantification result of the EDS was subtracted from the mean atomic number of the qBSE, resulting in a map of elements. Assuming Li is the only present element with Z < 4, the cipher calculation gives the Li content.

The SXES is equipped with two gratings, which enables analysis of two energy ranges of lower X-rays (JS50XL: 48–178 eV and JS200N: 53–211 eV). Especially the first grating allows the Li-K detection at 53 eV and Carbon-K at 138 eV. The measurements were performed at 2 kV with 40 nA and a dwell time of 100 s per point (point analysis and each point in a line-scan).

The colocalized Raman analysis was done using a Virsa Raman spectrometer. The inLux System is a retractable Raman probe and can be inserted between the SEM polepiece and the sample. The inLux can switch between a mirror for the video image, a Raman 523 nm laser and fiber to the Virsa spectrometer as well as a perforated mirror. This mirror containing a hole allows the direct view by the SEM. Calibrating the inLux video image and SE-image with a standard result in a colocalized analysis of features. The Raman spectra were recorded with 5 s duration at 10 mW laser power and 5 accumulations.

**ICP-OES** measurements allow the quantitative composition of the samples' metallic components to be determined. For this purpose, samples with 16 mm in diameter were dissolved in aqua regia. The remaining solids (such as graphite) were filtered off and the remaining solution was measured using ICP. The ICP-OES "iCAP Pro Series" (ThermoFisher Scientific) was used.

The measurement was calibrated using the ICP standard solution IV Certipur and the additional ICP standard for phosphorus from Sigma Aldrich. The following averaged

wavelengths were used for the evaluation: Li 610.366/670.784; Cu 324.754/327.396, Fe 238.204/259.94, Al 394.401/396.152, Co 228.616/238.892, Mn 257.61/260.569, Ni 216.556/221.647/231.604, P 177.495/178.285/213.618.

**Nuclear magnetic resonance** (NMR) and **electron paramagnetic resonance** (EPR) are complementary methods for analyzing materials that investigate the spins of the nuclei and the electrons, respectively. Both techniques have been used in the batteries field, ranging from investigations on the pure material properties such as structure and ion dynamics, over two-phase equilibria, to operando measurements of full batteries.[16,23–31] li Plating has been an important aspect in operando measurements,[32] as both techniques are sensitive to the bulk and able to distinguish between metallic, mossy, and dendritic Li.[17,32,33] Furthermore, the stage of Li intercalation into graphite is also reflected in the observed spectra. As long as metallic Li structures are thinner than the skin depth, NMR provides quantitative relative information.[16] However, absolute values of plated Li are generally not determined, and NMR has primarily been used to study cells with strong plating. With EPR, quantitative analysis of plating is more complicated since the signal amplitude depends on the density of states of the conduction electrons at the Fermi level, which may vary during battery cycling. Therefore, individual referencing is required for each studied system, for example by comparison with electrochemistry data.[32]

**NMR** spectroscopy was performed on a BrukerAvance III HD 400 MHz spectrometer equipped with a gradient probe (Bruker PA BBO 400W2/S4 BB-H&F-D-05 DIFF). For single-pulse 1D $^7$Li NMR spectra of solids, excitation pulses of 7.0 μs at 40 W with repetition delays of 10 s were used. Sample preparation took place in an Ar-filled glovebox. The anode active material was scratched off and filled into a Young-type NMR tube in powder form. These samples were used for solid-state measurements.

The **EPR** spectroscopy measurements at X-band frequency (9.323 GHz) were performed on a BRUKER ELEXSYS E580 spectrometer using a cylindrical TE011 mode resonator (SHQE) at room temperature. The spectra were recorded with 0.1 mW microwave power, 0.1 mT or 0.001 mT modulation amplitude, and 100 kHz modulation frequency. The EPR samples were punched out as 8 mm discs from the sheet in an Ar-filled glove box. The blanks are fixed between two half-shells of quartz glass in a 10 mm outer diameter quartz tube.

The electrodes used for **glow discharge optical emission spectroscopy** (GD-OES) and **neutron depth profiling** (NDP) analysis were unpacked in the glovebox under Ar atmosphere (O2 < 0.1 ppm, H2O < 0.1 ppm) and washed three times with dimethyl carbonate (DMC) for 1 min each to remove residual conducting salt. Three samples with a diameter of 16 mm were punched out from each anode from a central position, an edge position, and a corner position were taken for NDP. The samples dedicated for NDP measurements were fixed on adhesive Kapton film and covered airtight with 7.9 μm thick Kapton film. GD-OES samples were harvested next to the areas of NDP.

**GD-OES** samples were transferred from the glovebox to the GD-OES using an airtight transfer chamber under an Ar atmosphere. GD-OES analyses were conducted using a GDA750 device (Spectruma) on anode samples. The measurements were performed in radio frequency (RF) mode at a frequency of 2501 Hz, at a discharge voltage of 550 V, and a pressure of 2 hPa. A mixture of 1% H2 in Ar (both 6.0 purity) was used as a sputter gas. The following element-specific emission lines were used for detection: H (121.6 nm), O (130.2 nm), C (156.1 nm), P (178.3 nm), Si (288.1 nm), Li (670.7 nm, Ar (706.7 nm), and Ar (714.7 nm). The areas analyzed by GD-OES have diameters of 2.5 mm. The depth of the craters was determined after a measurement with a confocal microscope (Keyence VHX-7000). The Ar lines were detected with a charge-coupled device (CCD) camera, while the other lines were detected with photomultiplier tubes (PMTs). GD-OES depth profiling is a destructive ex situ method in which electrodes from disassembled Li-ion batteries are sputtered in an Ar plasma.[34–48] The sputtered sample atoms are then excited in the plasma to emit photons, which can be detected continuously, i.e., for each sputtering depth. A unique feature of the GD-OES method is that depth profiles ranging from the electrode surface to the current collector can be measured within a few hours. With an appropriate calibration, it is possible to quantify all elements relevant to Li-ion batteries, such as Li,[39,41] C,[41] O,[41] H,[41] P,[43] and Si.[42,43] In GD-OES depth-profiling measurements, the elemental content of Li can be measured, which may consist of different fractions: Li(GD-OES) = Li(metal) + Li(SEI) + Li(LixC6) + Li(Li silicates)

In previous studies on Li plating, anodes from aged cells discharged to the end-of-discharge voltage before disassembly were investigated, where the amount of metallic Li can be estimated under the assumption that nearly no Li is intercalated and the SEI consists fully of Li2O.[40,41,47] In the case of Si/graphite anodes, a part of the Li in the anode is bound as Li-silicates, which still allows us to estimate Li plating with an appropriate method.[40,47] In the present study, anodes from charged cells were investigated. Therefore, GD-OES is not able to distinguish between Li in $Li_xC_6$, Li in the SEI, and Li deposited on the anode surface.

**NDP** is a non-destructive nuclear analytical method, which has become a well-established tool for the quantification of Li depth concentration profiles in battery electrodes, thanks to its high sensitivity for $^6Li$.[49–52]

The NDP measurements were performed on the TNDP spectrometer of the NPI CANAM infrastructure at the nuclear research reactor LVR15 in Řež, Czech Republic (operated by the Research Center Řež). The TNDP spectrometer is placed on a short, vertically curved neutron guide, which filters a beam of the well-thermalized parallel neutrons with a flux of 6 × 10$^7$ n cm$^{-2}$s$^{-1}$.[53] During each measurement, the neutron beam irradiates the entire sample, which has a diameter of 16 mm. The samples have been punched out of two different graphite anode sheets (Sheet 2-anode, Sheet 10-anode) at SoC 100% at 3 different positions within the sheet (middle, corner, edge). The samples were protected with a Kapton foil (7.9 μm thick) to prevent Li oxidation on the anode surface. During the measurement, the sample is positioned toward the neutron beam at an inclination of about 5° to avoid neutron self-shielding. The 6Li(n,4He)3H nuclear reaction on the 6Li atoms in the sample is induced by the irradiation. The schematic principle of the reaction is shown in *Fig. 1*. The reaction products (alpha and triton particles) are isotropically emitted from the reaction sites and afterward detected by a fully depleted detector (a type of Canberra-Packard, F50143). The detector is positioned coaxially and parallel to the electrode surface at a distance of 55 mm. The charged particles gradually lose energy as they have to penetrate the electrode material, and afterward, they emerge at the sample surface. The energy loss depends on the path length, the material composition, and the material density. The Li concentration depth profile is obtained from the residual energy spectrum of the emerged particles.[54]

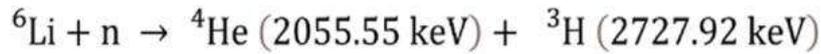

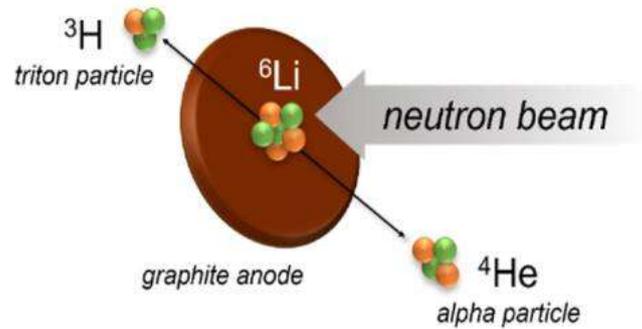

*Figure 1. The schematic principle of a 6Li(n,4He)3H nuclear reaction, which produces alpha (4He) and triton (3H) particles at well-defined energies.*

The energy spectrum of triton ($^3$H) is used to determine the Li depth concentration profiles within the electrodes because triton particles allow the detection of Li distribution in depths up to some tens of micrometers. In order to obtain sufficient statistics (of a few thousand counts per channel), the measurement time was 21–24 h per sample and depends on the Li content in the sample. The measurement of a reference standard with a well-defined $^{10}$B content was performed to determine the absolute amount of Li in the anodes. For both the energy calibration, i.e., to convert the detector channel numbers into an energy scale, and for the quantity calibration, a round-robin standard sample was measured. The content of 1.904 ± 0.015 × 10$^{16}$ $^{10}$B cm$^{-2}$ was determined after calibration to the SRM2137 standard.[55] For the determination of Li concentration in the anode samples, a natural abundance of 6Li (7.59%) is taken into account.[56] A constant average material density of 1.8630 g cm$^{-3}$ is assumed for each anode. The uncertainty in the final Li concentration measured by NDP is estimated to be 10%, and the depth resolution is 30 nm. The SRIM (Stopping and Range of Ions in Matter) software, which takes into account various layers of material with different compositions, is used to calculate the energy loss of tritons in graphite.[57] The data evaluation was performed by using the N4DP software.[54]

A complete overview of the used methods can be seen in **Table II.**

| Method | Measurement by | Spatial resolution | Quantification | Advantages/Challenges | References |
|---|---|---|---|---|---|
| Electrical Testing/DVA | ISEA | — | Yes | + Non invasive method | |
| | | | | + good correlation, between electrical data and Li-plating | |
| | | | | + no direct elemental information | |
| Flatbed scanner/Image Processing | ISEA | 100 μm–1mm | Colored based area quantification | + Easy, low cost, just colored based | 58$^-$61 |
| | | | | − No elemental information | |
| Light microscopy | HSAA | μm | No | + Fast method without sample preparation | 47, 62–64 |

| Method | Measurement by | Spatial resolution | Quantification | Advantages/Challenges | References |
|---|---|---|---|---|---|
| | | | | + color information | |
| | | | | − No depth information | |
| | | | | − no elemental information | |
| Laser scanning microscopy | ISEA | 100 nm–1 µm | Thickness | + Fast method without sample preparation | 58, 59 |
| | | | | + color, structure and height information | |
| | | | | + height profile of the depositions | |
| | | | | − No elemental information | |
| SEM/EDX/Raman/SXES | IAC | 0,7@20 kV–3 nm (SEM) | Yes (point, line, area) | + standard method for surface imaging and elemental analysis (SEM/EDX) | Cipher22: EDX65: |
| | | 0.5 µm (Raman) | | + colocalized analysis by using integrated equipment | |
| | | | | + Li detection (SXES) | |
| | | | | − Resolution depends on composition and accelerating voltage and current (SEM/EDX) | |
| | | | | − high energy impact can damage sample (SXES, Raman); limited spatial resolution (Raman) | |
| SEM/EDX | ISEA | nm | Yes (point, line, area) | + direct Li detection possible | 59, 60 |
| | | | | + fast sample preparation | |
| | | | | + high resolution | |
| | | | | − Small area | |
| FIB-SEM/EDX | GFE | 10–100 nm | Yes (point, line, area) | + high-resolution depth information with a possible EDX option | |
| | | | | − High preparation effort | |
| X-ray microscopy (XRM) | GFE/ISEA | µm | no | + 3D resolved structures | |
| | | | | − High preparation effort | |
| NMR | IEK-9 | 100 µm | Yes | + non-invasive | |

| Method | Measurement by | Spatial resolution | Quantification | Advantages/Challenges | References |
|---|---|---|---|---|---|
| EPR | IEK-9 | 10 μm | With referencing | − Quantification of Li difficult<br>+ non-invasive<br>+ Easily distinguishable between plating and charged electrode | |
| GD-OES | ZSW | 10–100 nm in depth | Yes | Depth limited to around 100 μm, allows detection of Li, destructive, challenging for lithiated anodes | 34–40 |
| NDP | MLZ/TUM | 10–100 nm in depth | Yes | Non invasive, depth limited to around 100 μm | |

## Results

In this section, the measurement results of all methods are presented individually. The summarizing discussion and evaluation of the results take place afterward.

For the detection of Li plating, the results of the methods used are presented, starting with the non-destructive tests, followed by the optical methods after cell opening, and then the spectroscopic methods.

The results of **the electrical cell cycling** are shown in *Fig. 2*. The aim was to force Li plating in a few cycles. *Figure 2a* shows the cell voltage curve for different current rates (C-rates) at +25 °C and 1 C charging at −5 °C. It can be seen that a higher C-rate, as well as a lower temperature, leads to higher overpotentials and, thus, a higher voltage level during charging. Since the cells were charged using a constant capacity charge, the maximum voltage is not predefined and can, therefore, be higher than the specified maximum cell voltage. For C/10, C/5, and C/3 at 25 °C, the maximum voltage is between 4.16–4.19 V and, therefore, only slightly above the recommended end-of-charge-voltage (EOCV) (see **Table I**). Charging with 1 C at 25 °C and at −5 °C leads to an end-of-charge-voltage of 4.27 V and 4.42 V respectively.

In *Fig. 2b*, the differential voltage analysis (DVA) results are shown. For this purpose, a C/20 discharge current was applied immediately after completing the constant capacity charging phase. The DVA curves at 25 °C look unremarkable, but with 1 C charging at −5 °C, the characteristic voltage plateau for Li stripping becomes visible. The end of the DVA plateau is at approx. $Q_{strip} = 0.4$ Ah, so that approx. 1.5% of the nominal capacity has been stripped. It should be noted that the Li stripping amount depends on the discharge rate, as Li is generally intercalated into the graphite at the same time as stripping takes place.[66]

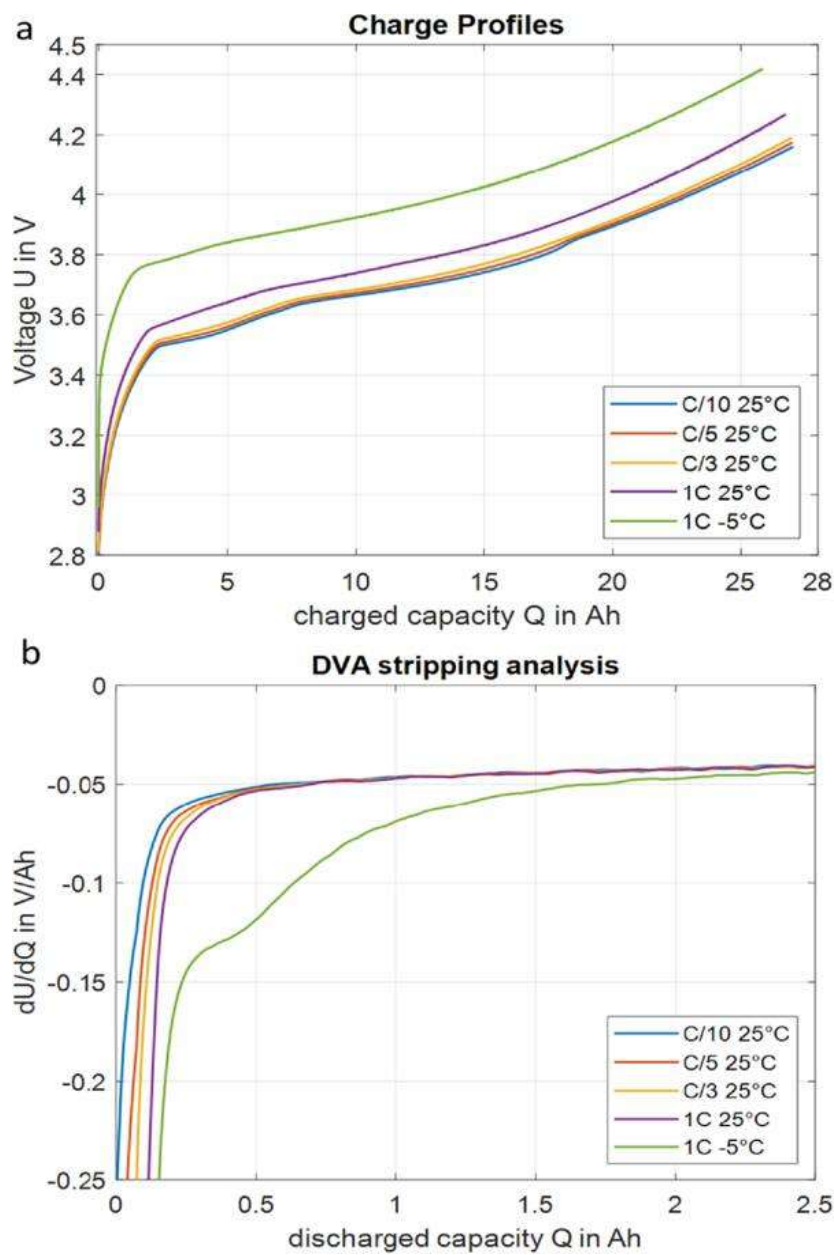

*Figure 2. Results of the electrical cell cycling. (a) Voltage curve during charging with different C-rates and temperatures. (b) differential voltage analysis (stripping test) with C/20 discharge current directly after fast charging.*

After the electrical tests, the cell was opened in a charged state (100% SoC) in order to examine the surfaces of the electrodes after charging. An uncharged cell was also opened and analyzed as a reference.

In the first step, the optical analyses were carried out directly after opening the cell.

*Figure 3* shows scans of the anode electrode surface using a **flatbed scanner** and the subsequent image analysis. In Fig. S4 an image of a discharged electrode can be seen for comparison. In *Fig. 3a*, an anode sheet from the stack middle and in *Fig. 3b* from the stack outside is shown. In both cases, fully lithiated golden $LiC_6$ is seen over most of the surface. At the edge of the electrode, a dark grey area that is not fully lithiated can be seen - the so-called anode overhang. The anode is geometrically slightly larger than the cathode due to, e.g., prevention of Li plating on the anode edge.[67,68] Therefore, an anode overhang can typically be found in Li-ion battery cells.

Nearby the overhang area, therefore close to the edge of the electrode, silver/grey/white depositions can be seen, which later turns out to be Li plating.

In addition to that, most of the depositions are next to the sheet edges; a preferred deposition edge can be observed across all sheets - the side of the cathode tab.

To quantitatively evaluate the scan results, both sides of the scanned anode electrodes were evaluated using image processing techniques. The results are shown in *Fig. 3c*. The anode surface of every sheet can thus be divided into 4 different areas. Overhang (cyan), charged graphite (blue), weak Li plating (red), and strong Li plating (yellow). In addition to sheet-level evaluation, a Li plating evaluation over the full cell stack is possible with a 3D representation (*Figs. 3d–3e*).

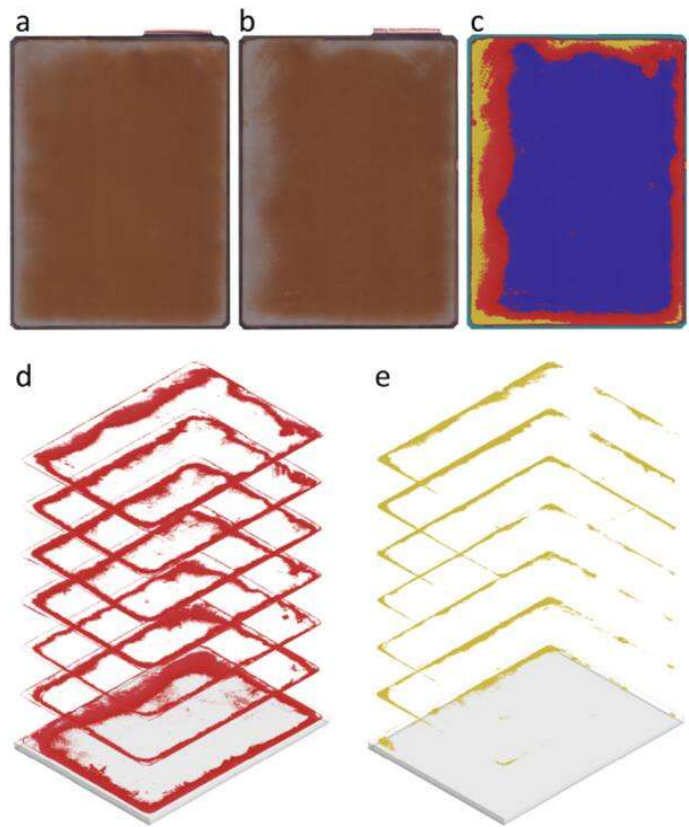

*Figure 3. Scanned electrode sheets after cell opening. (a) sheet from the stack middle and (b) from the stack outside. (c) image processing to evaluate the plated regions of the cell surface. Blue represents the non-plated areas (charged/golden graphite), red weak Li plating, and yellow strong Li plating. (d), (e) 3D-representation of the weak and strong Li plated regions over the full cell stack.*

For a better evaluation of the Li deposit structures, the magnification and resolution of the images was increased by using microscopes.

*Figure 4* shows the results of **light microscopy,** where a sample taken from the edge of the electrode was examined. *Figures 4a and 4a\** belong together and show an overview of the first 15 mm of the electrode from the electrode edge. In the edge region, a bluish-colored area of about 0.5 mm wide can be seen, which is the anode overhang. Blue indicates a low lithiation (presence of $LiC_{18}$). Further, towards the center follows a red-colored area ($LiC_{12}$), on which the first white-grey-colored Li plating can be seen.[69] The Li plating continues to increase until the entire surface is almost completely covered with Li deposits. Underneath the Li plating, a light golden coloration ($LiC_6$) can be seen. At a distance of 3 mm at the edge, there seems to be a local maximum of Li plating. Further, toward the center of the anode, the Li plating seems to decrease again. At approximately 7.5 mm (end of a and beginning of a*), the Li plating decreases significantly and appears only as an inhomogeneous haze-like layer on the surface. *Figures 4b–4d* show a higher magnification of the three color-marked areas. *Figure 4b* shows red and gold-colored graphite particles, which are covered with Li plating. On the one hand, the Li plating appears white and in a needle-like form called dendritic plating in the literature.[70,71] On the other hand, it appears insularly, and a clear microstructure cannot be recognized. Most of the Li plating in *Fig. 4c* shows a needle-shaped structure. Individual needles have a length of approx. 15 µm. Golden graphite particles can be seen under the Li plating. In *Fig. 4d*, the Li plating has already decreased significantly. The almost completely lithiated anode surface is only covered by individual Li plating needles.

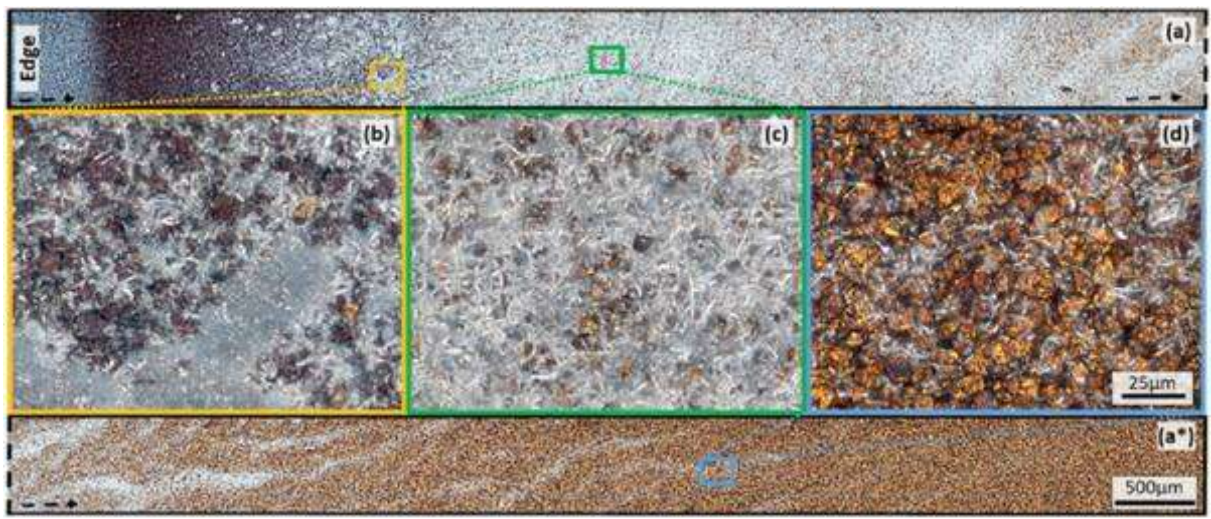

*Figure 4. Light Microscopy Results. (a) and (a\*): overview of the first 15 mm of the electrode (b)–(d): higher magnification at marked positions of a (same scale as in image d).*

A higher resolution was achieved with a **laser scanning microscope**. This device also provides height information about the sample. An area with clearly visible macroscopic deposits at the cell edge was analyzed in more detail in *Fig. 5*. In both magnifications (*Figs. 5a–5b*), on the one hand, golden/red colored graphite particles can be identified, which are covered by silver-colored needles with a length of approx. 10 µm and a diameter of a few micrometers. Such needle-like structures are a frequently reported morphology of Li plating in the literature, and it is called dendritic plating.[72] *Figure 6c* shows a height profile of the anode electrode surface. It can be clearly identified that the needles are located on top of the graphite particles, thus growing towards the separator and leading to an increase in cell thickness and, in the worst case, to an internal cell short circuit.[73]

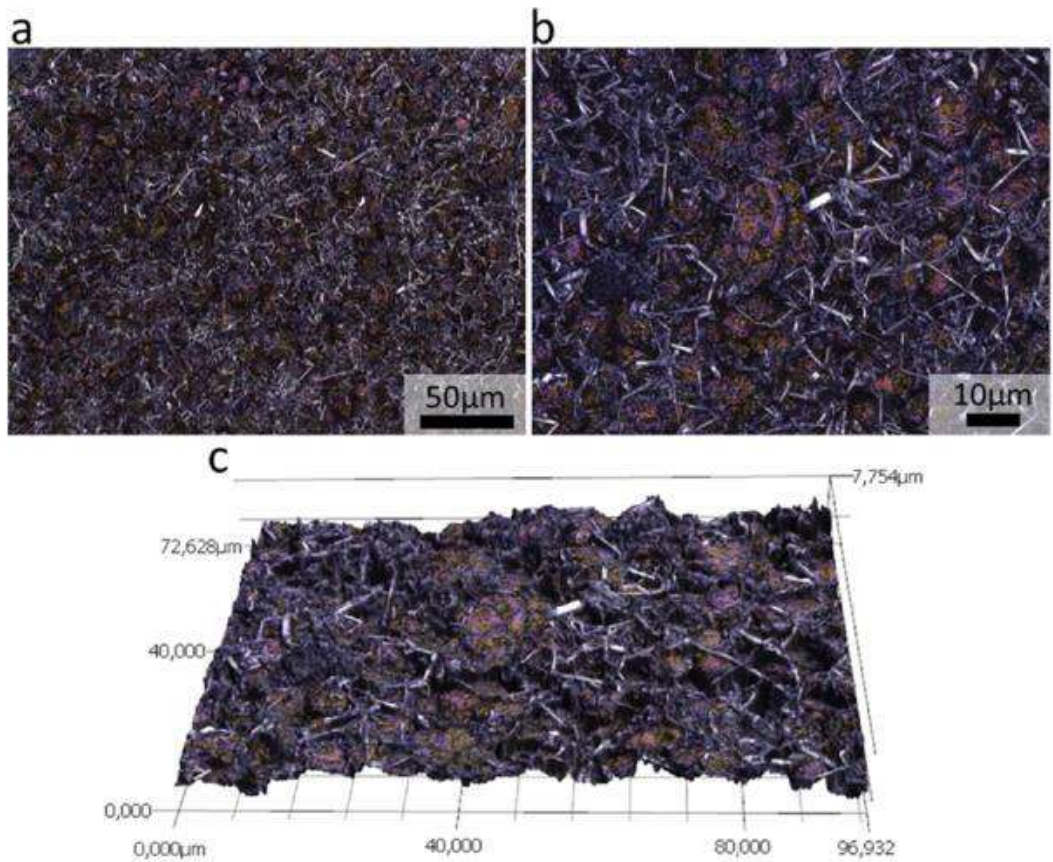

*Figure 5. Anode surface from the plated area imaged by laser scanning microscopy. (a) 50× magnification. (b) 150× magnification. (c) 3D-height profile of image (b).*

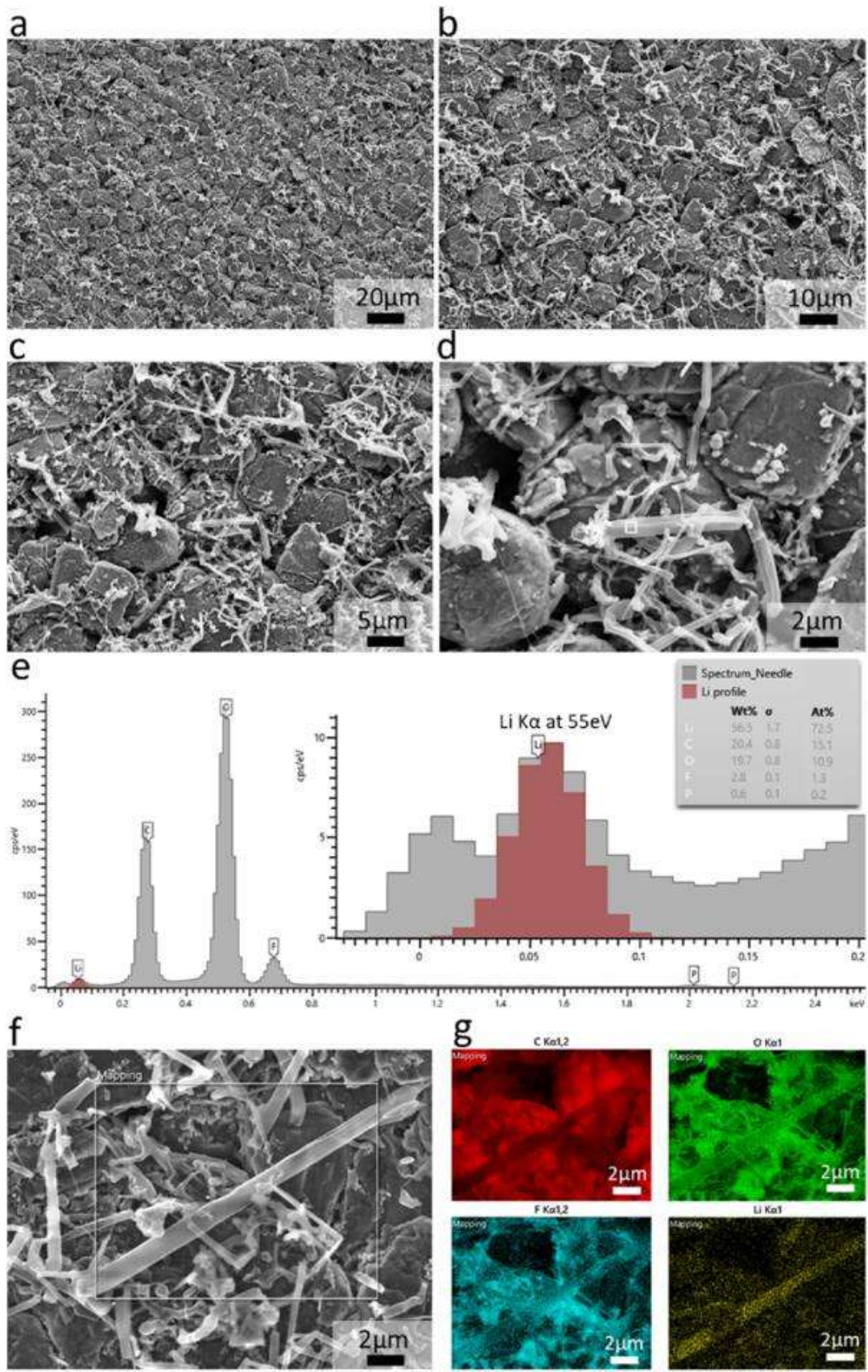

*Figure 6. (a)–(d) Anode surface from the plated area imaged by scanning electron scanning microscopy at different magnifications. e EDX-spectrum of a spot measurement directly on a needle-like structure. f-g EDX mapping of an area with plated structures.*

**SEM** measurements were carried out to confirm the results and further increase the resolution. In *Fig. 6*, the results of the **SEM/EDX** evaluation are shown. In *Figs. 6a–6d*, the anode surface is shown for 4 different magnifications. At the higher magnifications, needle-like structures are clearly visible with the same length/diameter (10 μm/1 μm) as also previously seen in the light/laser microscopy. In addition to the morphological surface image, an EDX spectrum of a needle-like structure was measured (*Fig. 6e*). In this spectrum, a clear Li Kα-peak at approx. 55 eV can be observed. It is important that the Li peak can be clearly distinguished from the noise peak at 0 eV. With standard detectors (i.e., with a beryllium glass window), the noise peak is so wide that the Li peak is not visible. The calculated weight percentages and atomic percentages result in 56.5 Wt% and 72.5 At%, respectively. The statistical error is displayed as σ for the calculated wt% in the image legend.

In addition to point measurements, an EDX mapping of a deposited area was recorded (*Figs. 6f–6g*). The most frequently occurring elements are carbon, oxygen, fluorine, and phosphorus. Carbon is the main element of graphite and part of the organic electrolyte, just like oxygen. Fluorine and phosphorus contents are expected from the $LiPF_6$ conducting salt. In the mapping, the dendritic structures can be displayed by their Li content. The carbon image is exactly the opposite in the spatially resolved concentration.

An area with obvious Li plating (silver, needles) was chosen for the colocalized analysis in the JEOL IT-800. The SEM image, as well as the respective elemental maps, are shown in *Fig. 7a*. Measurements were conducted at accelerating voltages of 5 kV and 15 kV, with currents adjusted to enhance signal output at the lower voltage. Using different voltages allowed us to gather information from near the surface at 5 kV, while the mapping at 15 kV captured signals from deeper within the material. Consistent with previous findings, EDX identified carbon, oxygen, fluorine, and phosphorus (details in the SI). After applying the Cipher procedure, the elemental maps were rescaled, resulting in the addition of a Li map. The main structural features of the SEM image were reproduced, with the mapping at 5 kV appearing more detailed. The mappings clearly indicate an overall medium to high Li content, except for a particle in the top left corner, which is predominantly carbon. The intensity of carbon signals is higher at 15 kV compared to 5 kV, likely due to the inhomogeneous distribution of Li in the carbon matrix, suggesting that Li deposition predominantly occurs at the surface.

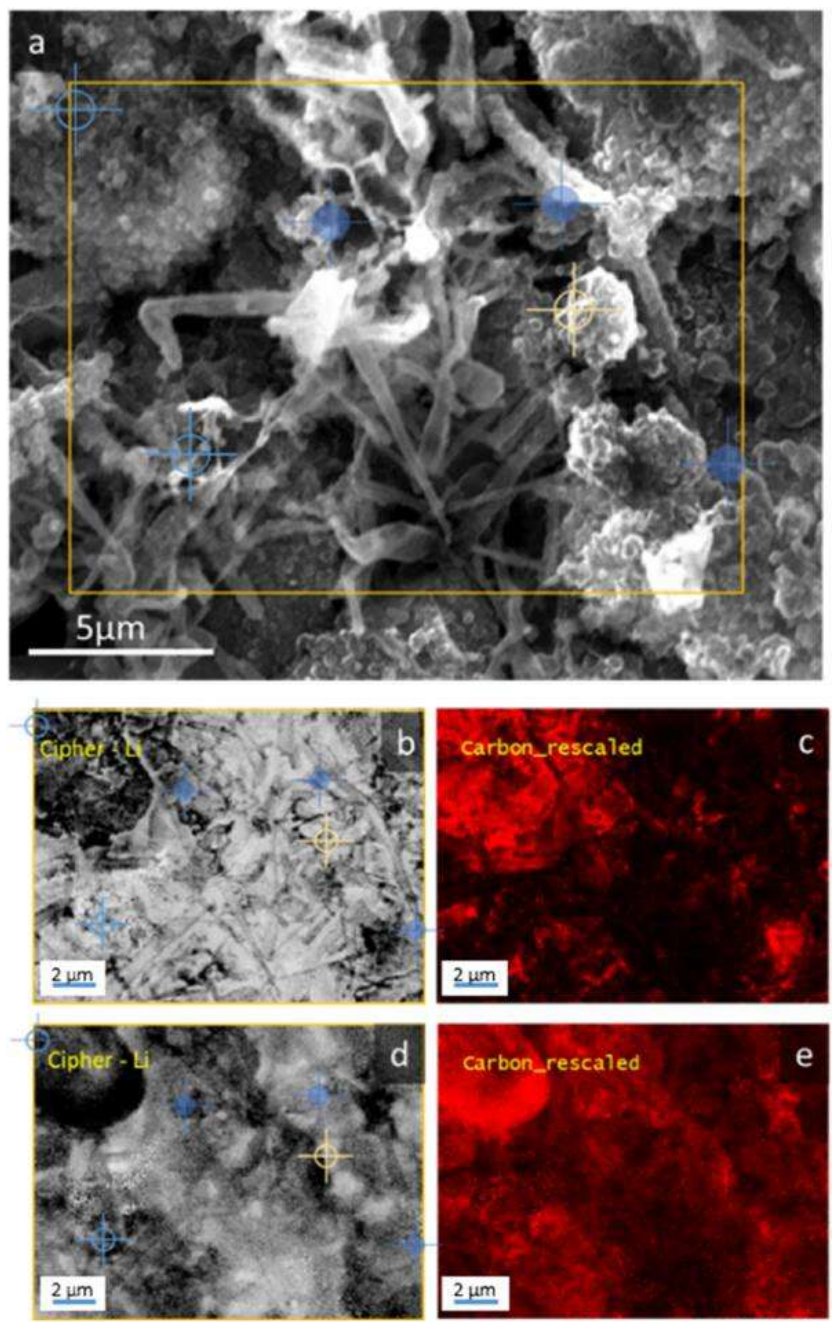

*Figure 7. (a) SEM image of an investigated area by Cipher, SXES, and Raman. The yellow frame depicts the mapping for EDX with its respective calculated maps by Cipher (b)–(d). Spots marked in blue denote the investigation by SXES, whereby the presence of Li is represented by filled circles. The spot, which was measured by Raman, is colored in yellow. EDX-maps taken at 5 kV, 500 pA (b), (c) and 15 kV, 115 pA (d), (e) after Cipher procedure showing the Li as well as carbon distribution.*

SXES spot measurements were conducted to verify these results. The spectra showing transitions of Li (Kα1), C (Kα2 and Kα3), O (Kα3 and Kα4), and F (Kα1, Kα2, and Kα4) with varying intensities are available in the SI (Fig. S1). The presence of Li, indicated by the signal at 54 eV (Li Kα1), is marked by filled blue circles in Fig. 7 for stronger signals and by open blue circles for weaker signals.

Due to the low acceleration voltage of 2 kV used for recording SXES spectra, the information obtained is predominantly from the surface of the material, correlating well with the Cipher maps taken at 5 kV. As additional validation, Raman spectroscopy was performed on the spot (marked in yellow in *Fig. 7*) within the area mapped by Cipher. Because of the energy impact from both the electron beam for SXES and the laser for Raman, spectra were not taken at the exact same spots. Despite the high surface roughness, which causes significant scattering and is not ideal for Raman spectroscopy, the analysis is expected to reveal the D-band and G-band at 1319 cm$^{-1}$ and 1586 cm$^{-1}$ for graphite. These signals are known to shift slightly to higher wavenumbers, around 1363 cm$^{-1}$ and 1615–1640 cm$^{-1}$, for lithiated graphite.[74]

At full lithiation, the D- and G-bands disappear. During overlithiation, a signal around 1825 cm$^{-1}$ becomes visible, attributed to Li$_2$C$_2$ formed in situ by the laser. In this study, the investigated spot (marked in yellow) shows a spectrum (Fig. S2) with overlapping D- and G-bands. A curve fit reveals a D-band at 1369 cm$^{-1}$, which is close to the reported value in the literature (Table S1). However, the G-band is unexpectedly shifted to 1532 cm$^{-1}$. Despite this change, the position of the D-band indicates the presence of lithiated graphite. It should be noted that the Raman spot size, approximately 1 μm, is relatively large and may be influenced by surface structure.

Further measurements on two different sample areas (*Fig. S3*) showed shifted D- and G-Bands in both cases. One measurement was taken on a graphite flake, revealing only shifted signals for D- and G- Bands. The second area, which contained needle-like structures similar to those in *Fig. 7*, exhibited an additional strong signal at 1875 cm$^{-1}$. This signal indirectly indicates the presence of deposited Li.

The SEM measurements provided a view of the top of the sample. To get an impression of the inside of the electrode, the sample was cut using a FIB (*Fig. 8*). In combination with the installed EDX sensor, it was possible to analyze the materials. Carbon was found in the deep areas. Oxygen, fluorine, and phosphorus were detected at the top of the electrode. The image also shows a different structure of the top layer, which does not look like the underlying carbon particles.

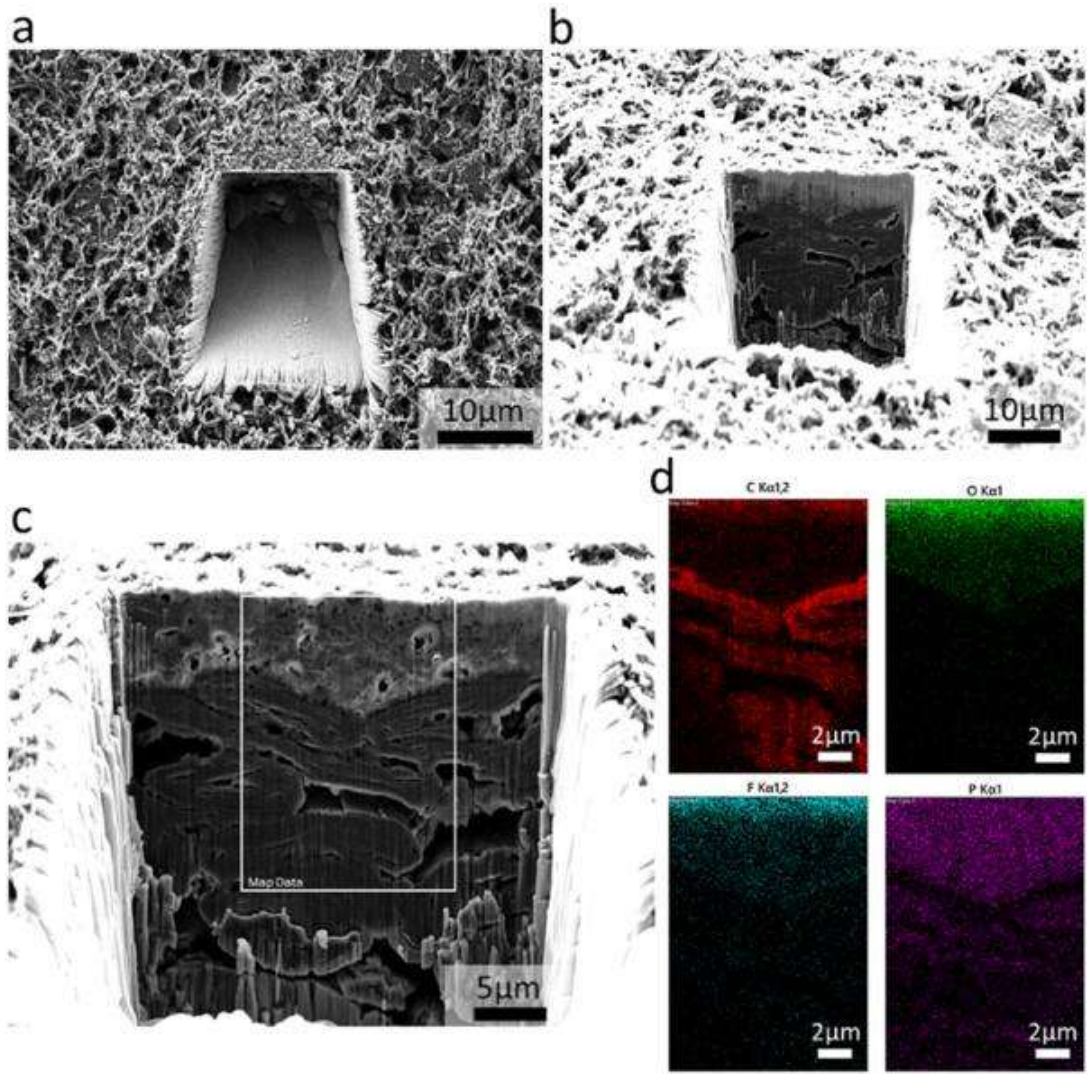

*Figure 8. Results FIB-SEM/EDX. (a) View of the electrode surface around the FIB cut (b) and (c) depth view; (d) EDX mapping of the marked area in c).*

X-Ray Microscopy allows an inside view of the electrodes (*Fig. 9*). The electrode structure and particle size distribution can be visualized. The images show a different material structure in the inner electrode and the outer edge (Figs. 9c and 9d). The structural change allows the assumption that deposits have formed on the surface. The material itself cannot be identified. The sample's structural resolution allows the thickness of the differently structured layers to be determined.

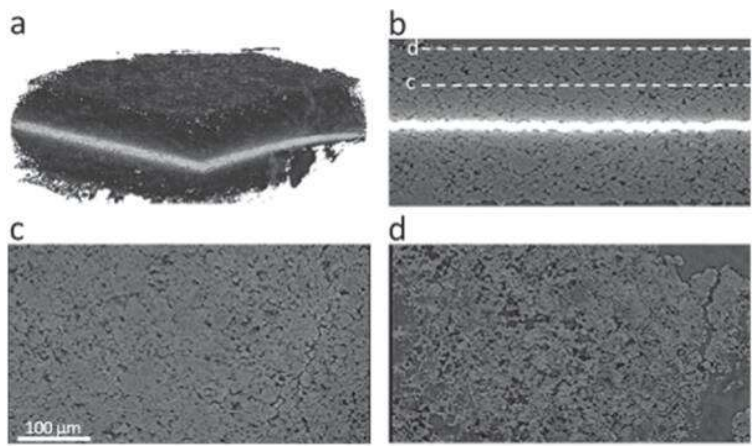

*Figure 9. Results XRM (a) Anode 3D view, (b) cross-section image including two lines for the positioning of images c and d, (c) plane view of the middle of the electrode, (d) plane view of the upper area including depositions.*

The optical results shown so far provide information about the location and structure of the deposits on the electrode. The results of the spectroscopic EDX method have already been presented and are now substantiated by further spectroscopic methods. Further **spectroscopic** investigations will prove that the visible deposits can also be chemically assigned to Li metal.

To qualitatively evaluate the results of the various methods, ICP-OES measurements are also carried out to determine the amount of Li in the total material. The current collector thickness can be calculated based on the amount of copper and aluminum in the samples (**Table III**). The discharged cell already shows Li in the anode sample due to the SEI layer being out of the formation process and not completely delithiated graphite at the end of the discharge voltage. For the charged cell, anode samples of plated (shiny silver) and not plated (gold) regions were analyzed. The results show nearly the same amount of Li in the samples.

|  | **Current collector thickness** | **Li in μmol/cm$^2$** | **Plated area Li in μmol/cm$^2$** |
| --- | --- | --- | --- |
| discharged cell cathode | 10 μm | 173 |  |
| discharged cell anode | 20 μm | 20 |  |
| charged cell cathode | 10 μm | 89 |  |
| charged cell anode | 20 μm | 98 | 100 |

**NMR** allows the identification of the materials via the excitation of individual nuclides and the evaluation of their environment depending on the specific responses. Three different samples were measured: a fully charged anode (golden-colored sample), a discharged sample (grey-colored sample) and a sample in which Li plating was assumed (shiny silver-colored area). Solid-state $^7$Li spectra of all three materials are shown in *Fig. 10*. The signal maxima at 42.9 ppm, along with the characteristic non-central quadrupole transitions (with maxima at about 188, 109, −32, and −94 ppm) from the charged and Li plating assumed sample spectra, are scaled to the same value. This allows a semi-quantitative comparison of these two samples. A close-up is shown in *Fig. 12*. Their spectra show largely identical features, differing only in the regions around 0 ppm (marked in light blue) and an additional phase-shifted signal around 265 ppm for the silver-colored sample. The range between about 25 ppm and 70 ppm corresponds to $LiC_6$ and $LiC_{12}$. While it cannot be further resolved here, the exact same position of the maximum indicates a very similar degree of intercalation,[16] justifying the signal scaling. Signals in the high-ppm region (in this case 265 ppm) have been associated with plated metallic Li.[75] This signal amounts to about 0.3% of the total signal area. The signal surplus of the shiny silver-colored sample around 0 ppm accounts for about 3% of the total signal area.

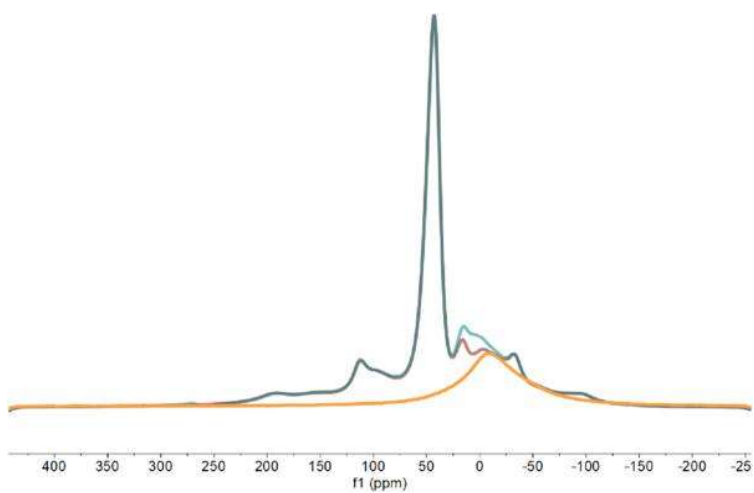

Figure 10. $^7Li$ overview spectra of a charged anode (red) and a plating assumed sample (blue) normalized with respect to their signal maximum and referenced to the signal of $LiC_6$ at 42.9 ppm. The discharged anode (orange) is externally referenced to aqueous LiCl.

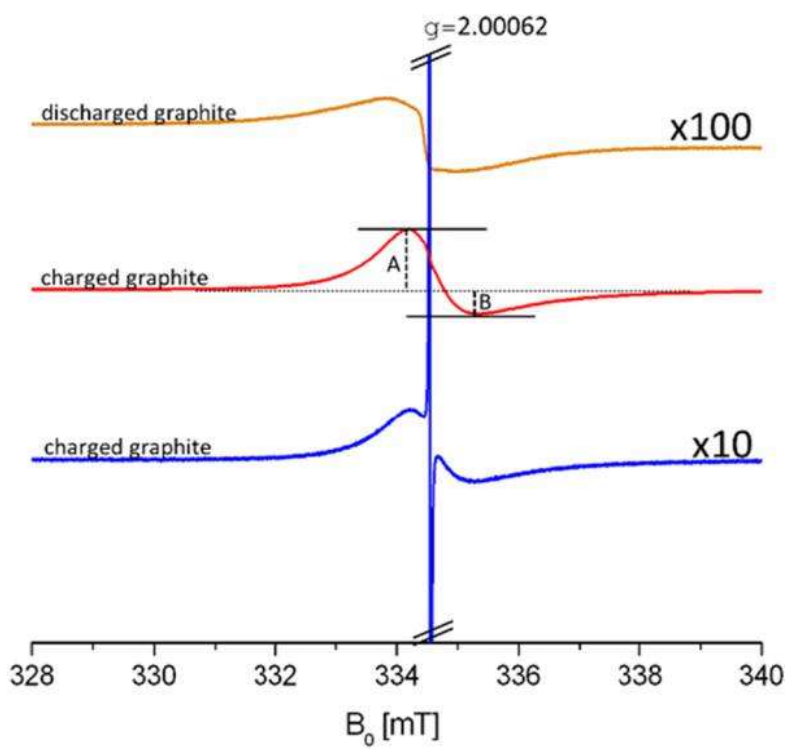

Figure 11. EPR spectra of a discharged graphite sample (orange), a charged one (red) and a silver one with metallic deposits (blue).

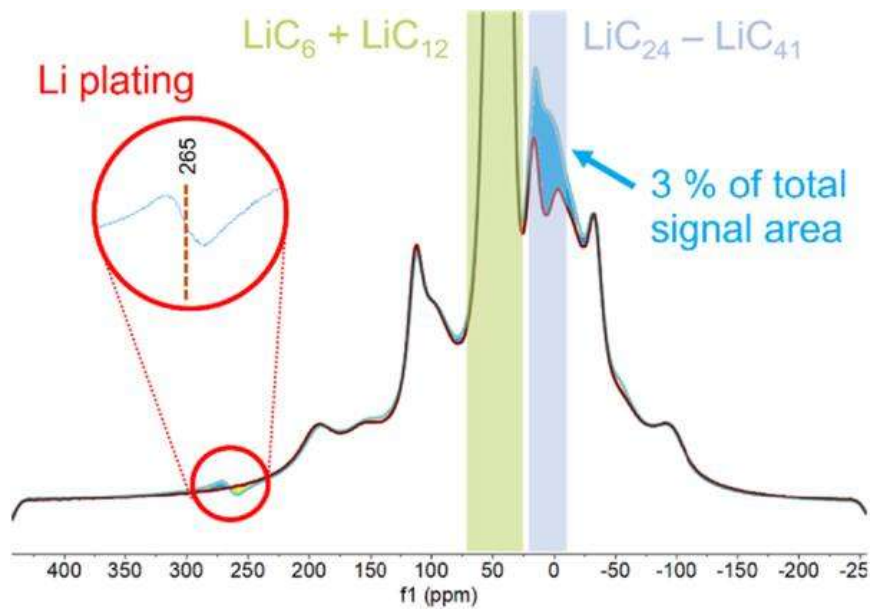

*Figure 12. ⁷Li close-up spectra of a charged sample (red), and the silver-colored sample (blue).*

Different origins for signals in the 0 ppm region have been reported. $^7$Li resonances for low amounts of intercalated Li (LiC$_x$ with x>12) in graphite were found at around 10 ppm. For LiC$_{18}$, a chemical shift of about 13 ppm has been reported,[16] with slightly upfield shifted resonances for lower intercalation levels.[76] At about 0 ppm, lithium salts are found.[77] Furthermore, several authors have reported a signal between 0 and 10 ppm that occurs in fully intercalated or overintercalated graphite.[75,78,79] A closer look at the difference between the spectra of the charged and potentially plated samples indicates that multiple additional contributions may be present in the latter.

In particular, the potentially plated sample contains a signal component that is similar to a resonance reported for an overintercalated highly oriented pyrolytic graphite (HOPG) sample. After a long resting period of several months and further heating of the sample, a high-ppm signal (~270 ppm) evolved towards a signal around 10 ppm, which has been associated with local overintercalation of small amounts of Li beyond LiC$_6$.[79] The reason why the resonance between 0 and 10 ppm is more intensive than the resonance at 265 ppm in the silver-colored sample may be indicative of dynamic processes occurring after the end of the cold charge cycle, yet this needs further investigation in the future.

The spectrum of the discharged sample, does not show any of the characteristics that are typical for strongly intercalated materials. Only a broad signal without fine structure is observed with a maximum at −8.5 ppm, which is characteristic for non-mobile Li in an electrically non-conductive environment, such as an SEI.[75] This feature is very different from the signal difference between the charged and assumed plated samples, indicating that the latter is actually correlated with Li involved in plating.

EPR measurements are complementary to NMR This method takes advantage of the higher gyromagnetic ratio g of electrons compared to nuclei and, thus, a significantly higher sensitivity compared to NMR when unpaired electrons are present. In EPR, different morphologies of metallic Li can be distinguished by the line shape.[17]

The signal of delithiated graphite (*Fig. 11*) is composed of two components. A broad signal (g = 2.00147) with a peak-to-peak linewidth $\Delta B_{pp}$ = 1.2 mT (Gaussian) and a narrow signal

(g = 2.00124) with $\Delta B_{pp}$ = 0.26 mT (Lorentzian). The narrow signal may originate from free spin carriers and localized unpaired electrons[80] probably from the carbon black. The broad signal originates from the graphite.

The intercalated graphite in the charged sample gives the well-known asymmetric Dysonian EPR signal. It can be characterized by $\Delta B$ = 1.1mT and an A/B = 2.4 ratio (see *Fig. 11* for definition of A and B), which can be attributed to the metallic electrical conductivity and skin effect.[81] This Dysonian line shape is characteristic for conductive samples with dimensions larger than the skin depth. For samples with smaller dimensions, such as Li coatings, a Lorentzian or near-Lorentzian line shape results.[82]

The graphite in the silver-colored sample shows deposited metallic Li with a very narrow signal ($\Delta B_{pp}$ = 0.003 mT) at g = 2.00062. Therefore the formation of mossy Li can be detected very sensitively. The $LiC_6$ signal shape remains the same compared to the non-plated, charged sample. However, the intensity is 10 times smaller. This could be explained by a higher electrical conductivity of the sample or by stronger shielding effects due to a high local electrical conductivity. Multi-layered systems that present an impedance mismatch to the irradiated electromagnetic wave at their interface lead to reflections, whereby the signal of the underneath layer is attenuated disproportionately compared to the skin effect. Therefore, quantitatively may be approximately achieved only for the top layer, but not for layers below.[83]

To verify the data from the optical FIB measurements methods were investigated that offer a depth resolution.

*Figure 13* shows **GD-OES** depth profiles from anodes taken from cells disassembled in the charged state. The bulk value corresponds to Li in $Li_xC_6$. However, a high near-surface Li peak (0 μm < depth < 10 μm) is detected for charged electrodes with GD-OES, which could indicate both Li deposition and a higher concentration of Li intercalated into graphite on the surface.

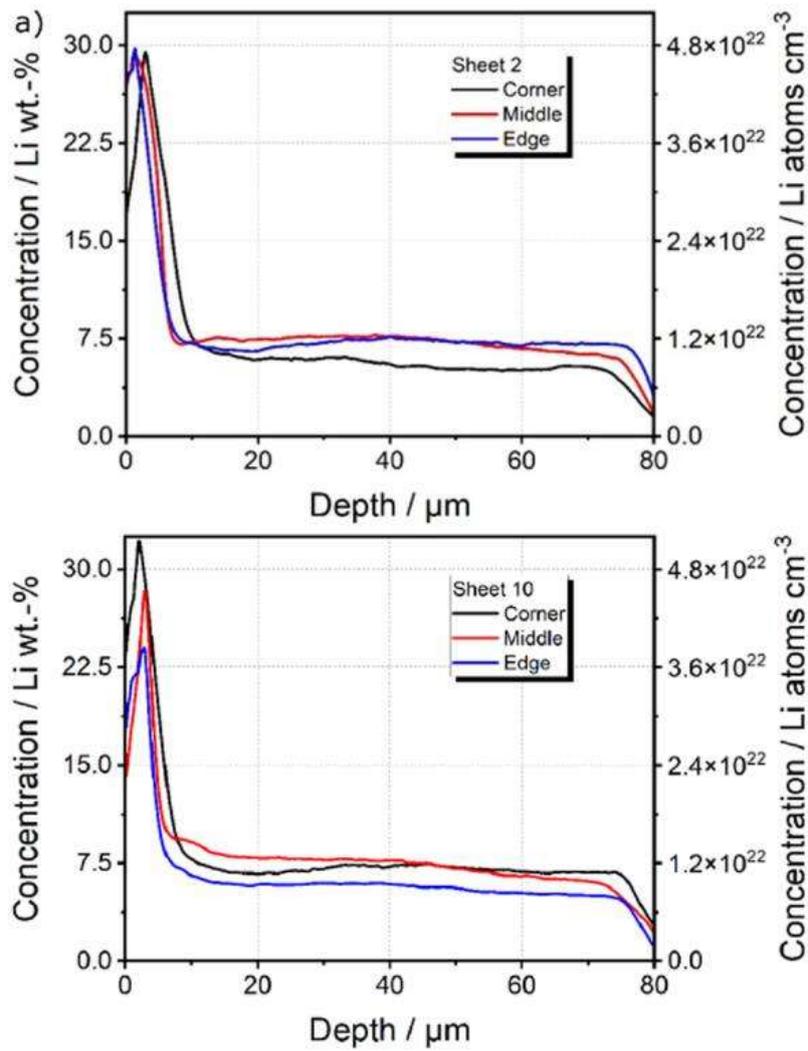

*Figure 13. Ex situ GD-OES depth profile of the anode from the aged cell investigated in the present study (a) sheet 2, beginning of the electrode stack, (b) sheet 10, middle of the electrode stack.*

*Figure 14* shows the Li concentration depth profiles obtained by **NDP** for two different types of graphite anodes at the SoC 100%. Sheet 2-anode has the highest Li concentration of 1.58 × $10^{22}$ li atoms $cm^{-3}$ measured in the corner of the anode with a maximum depth of the Li plating of 2–4 μm. This finding is attributed to the inhomogeneous Li plating, which is most pronounced in the corners (highest Li concentration). The maximum concentration measured on the edge is 1.33 × $10^{22}$ li atoms $cm^{-3}$ and in the middle is 9.36 × $10^{21}$ li atoms $cm^{-3}$. The Li concentration becomes equal in the bulk after the depth of ~8 μm for all three positions. The Li concentration of Sheet 10-anode results in the highest Li concentration being 1.21 × $10^{22}$ li atoms $cm^{-3}$, which is the same in the corner and on the edge part of the electrode. The highest concentration measured in the middle is 9.90 × $10^{21}$ li atoms $cm^{-3}$. The Li concentration becomes equal in the bulk after the depth of ≈ 4 μm for all three positions. The highest Li concentration in Sheet 10-anode is lower than the highest concentration in Sheet 2-anode.

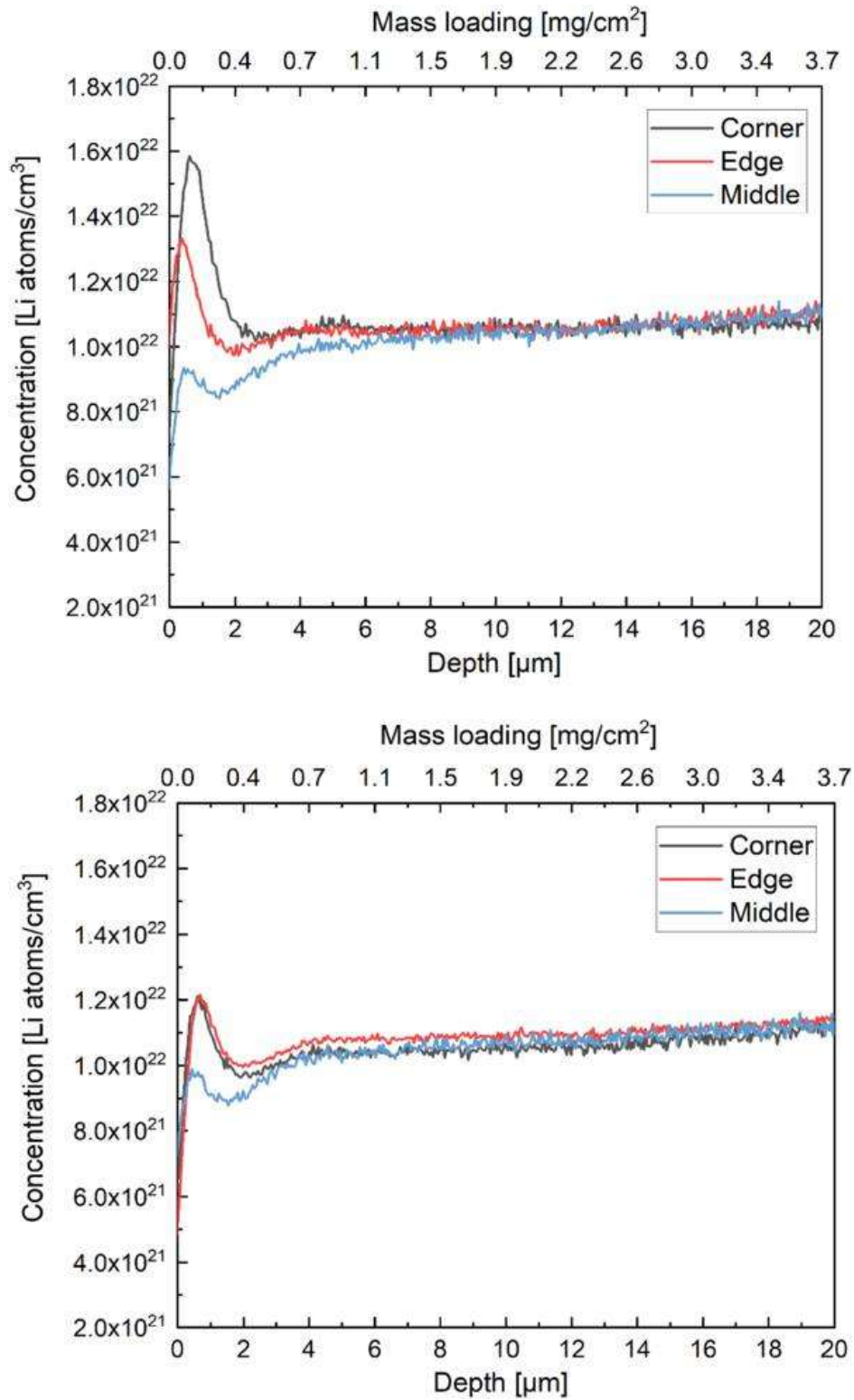

*Figure 14. (top) Li concentration profile vs depth on three different positions (corner, edge and middle) of Sheet 2-anode (outer part of the electrode stack) and (bottom) of Sheet 10-anode (middle part of the electrode stack).*

**Discussion**

After the results are presented, the findings are linked, discussed, and evaluated. Here, the methods are categorized, their possibilities are highlighted, and limitations regarding the detection of Li plating are discussed.

The images taken (scanner, light microscopy, laser scanning microscopy, SEM) show deposits and shiny metallic areas. The deposits' structures are needle-like, as can be clearly seen from the high-resolution methods. The optical methods combined with the EDX results allow the identification of metallic Li on the surface of a sample.

The SEM-EDX analyses up to Li were used to identify the metal of the needle-shaped deposits. The EDX measurement shows that over 70at% metallic Li is present. Furthermore, graphite can be identified as the basic substrate. It is assumed that the fluorine and phosphorus source is the $LiPF_6$ conducting salt or degradation products thereof. The samples were not washed beforehand so that conductive salt residues may be present. The measured elements may result in irreversible and electrically insulating lithium products (e.g., $Li_2CO_3$, $Li_3PO_4$, or LiF), which then become electrically inactive and passivate the surface.[84–86]

The windowless detector allows the detection of (metallic) Li. This detection is improved if no further reaction of the Li has occurred beforehand,[87] so ideally, the cell is opened as quickly as possible, and the samples are transferred to the SEM in the absence of an ambient atmosphere. By detecting the needle-shaped deposits as Li using the EDX detector, the shiny metallic areas with needle-shaped structures can be reliably assigned to Li metal deposits. In the long term, this method makes it possible to fall back on cheaper and faster microscopy images and only have to carry out a few measurements with the SEM-EDX for verification.

The FIB-SEM-EDX analyses confirm the assumption that graphite is the basic substrate at depth. The depth section shows that a layer has formed on the graphite electrode, which cannot be identified by the standard EDX detector. In combination with the already described results of the windowless EDX detector, it can be confirmed, that the not detectable element must be Li. The advantage of FIB measurements is that you can see the depth profile through the generated cross-section. Cutting the sample with the ion beam does not definitively avoid a change in the sample. However, in combination with the other methods, a strong change can be discounted. The XRM measurement counteracts this disadvantage of the destructive FIB measurement method.

XRM images were taken to get an insight view of the sample. Due to the resolution and the material density, it is challenging to detect metallic Li properly. The images show differences in the material structure. Therefore, these images can be used to confirm that the sample has different material structures and the positioning of the different layers. Due to the different radiopacity, it is possible to allocate the materials to dense and lighter materials.

The combination of SEM-EDX and Raman allows the uncharged particles to be viewed and evaluated in detail. The focus is on the uncharged particles, as these are noticeable due to the overall charged state of the sample (100% SoC). In SEM-EDX, all carbon-based particles are labeled in the same way. Using the additional Raman method, these particles can be separated into graphite-based particles, which can no longer be charged due to, for example, a loss of contact, and non-graphitized carbon particles as Particle II in Fig. 7, which are present in the active material. This step allows a more detailed analysis of the active materials in terms of cell aging and active material loss. In the past, uncharged particles in an overall charged electrode were automatically categorized as active material loss. With this method, it is possible to include the composition of the active material in the analysis.

The NMR results show just minor differences in the lithium signal of a plated and a lithiated electrode.

The discharged sample gives a broad, featureless $^7$Li NMR signal with its maximum at −8.5 ppm. No contributions of the former samples' signal are observed. Hence, Li is uniformly removed from the electrode material, and apparently, no clusters with higher Li concentrations remain. The peak's intensity (*Fig. 10*) suggests that up to 30% of the original Li is still in or on the anode material after delithiation. However, a quantitative comparison of the delithiated material with the intercalated materials is questionable. The later ones show metallic properties,[88] hence, a limited penetration depth of the radio frequency pulse in the material is expected, leading to a non-uniform excitation of Li in the material. For graphite, the conductivity is different, leading to an altered penetration depth and a higher volume of the material contributing to the signal, which results in a potential overestimation of the remaining Li. For benchmarking of the NMR based quantification, the ICP-OES results can be used. These results show that in the discharged state 10% of the overall Li is still in the anode due to SEI formation. In the case of the fully charged cell, around 50% of the Li is in the anode. Therefore, from the cyclable amount of Li, approximately 20% are bound in the anode in the discharged state. The assumption that NMR overestimates the amount of Li in the discharged electrode is thereby confirmed, yet results of NMR and ICP-OES are of comparable magnitude. A more quantitative NMR analysis may be possible by using an internal reference such as $^{13}$C NMR from graphite for normalization.

The NMR signal integral ratio of silver (plated) to gold samples, with an upper limit of 1.03:1, are consistent with ICP-OES results, where plated samples showed approximately 2% more lithium per cm$^2$ in the anode.

Fang et al. have previously reported the combination of NMR and ICP-OES for quantitative distinction between plated Li and Li in an SEI.[75] However, they investigated cells post-test after substantial plating had taken place, leading to a significant signal at around 270 ppm. Here, we investigate the onset of plating, which shows different spectral features. Furthermore, our results indicate that care must be taken regarding the integration of parts of phased spectra, since the high-ppm metallic Li signal shows a different phase than the signal from intercalated Li, which may lead to systematic errors if not considered in the data analysis. While the NMR measurements showed semi-quantitative signal changes upon Li plating, the sensitivity was rather low, requiring a high number of scans and, therefore, a long experiment time to detect the small amount of Li plating in industrial battery cells. EPR, on the other hand, requires extensive calibration measurement for the quantitative detection of Li plating but is able to detect the onset of Li plating much more sensitively.

Furthermore, EPR allows the detection of different Li structures and morphologies in a sample. The discharged sample shows a signal consisting of two components. The dominant broad part of the signal may originate from the graphite itself. But increasing amounts of Li in graphite can be distinguished. With fully intercalated graphite, a signal with similar width, but a Dysonian line shape is observed, and a narrow signal is visible in case of plated Li.

An interesting aspect of this work is the EPR line width caused by the plated Li. In test cells, where quantitative plating had been induced, such narrow lines had been observed for dendritic Li growth, while thicker layers of porous, mossy Li caused a somewhat broader line with $\Delta B_{pp} = 0.03$ mT.[17] Therefore, the line width is not simply a function of the Li morphology, but rather a complex function of morphology and layer thickness. For commercial cells, where plating is an unwanted side effect, the observation of such narrow lines increases the sensitivity of EPR further to detect the onset of plating. The discharged and fully charged samples do not show a sign of the onset of metallic Li formation.

The EPR measurements confirm the EDX measurements that plated Li is present in the silver sample. The EDX measurement only allows a statement to be made about the surface, whereas the EPR measurement considers the entire material thickness and also provides information on potential metallic Li nucleation inside the graphite. On the other hand, EDX measurements allow a spatially resolved statement to be made by selecting a specific area or point in the SEM image. With EPR imaging a similar spatially resolved investigation may be possible at a lower resolution.[17] However, EPR imaging protocols do not currently allow for an artifact-free distinction between signals with different phases, i.e. symmetric and Dysonian lines at the same time.[89]

The ICP-OES results can be used to determine the total amount of Li present in the sample. This measurement technique does not allow spatially resolved analysis within a sample. Accordingly, no Li plating can be detected with the ICP-OES method, as no distinction can be made between intercalated and metallic Li.

After analyzing the surface (optical and EDX) and the bulk material using EPR, NMR, and ICP-OES, the depth distribution will be investigated. Optically, this includes the FIB images and, in spectroscopy, the GD-OES and NDP measurements.

The Li contents in the bulk (depth > 10 μm) in the GD-OES measurements agree well with the values determined by the NDP (compare *Figs. 13* and *14*).

The peak (*Fig. 13*) at the surface (0 to 10 μm depth) may originate from different sources such as i) preferential sputtering, ii) SEI growth, and iii) inhomogeneous lithiation of graphite. Preferential sputtering, which was recently demonstrated for the depth profiling methods GD-OES and NDP,[90] could result in very high detected surface concentrations in GD-OES. In addition, the SEI is also typically most pronounced at the anode surface.[48] For anodes investigated in the present study, the SEI accounts for ~21.7% of the detected Li. Quantitative determination of $Li_xC_6$ is not possible with GD-OES. These results demonstrate the current limitations of the GD-OES method for charged electrodes. Further studies are required to unravel the causes of the Li peak and to develop a method that can be used to reliably measure charged electrodes. Li depositions are more easily detectable by GD-OES on anode samples from discharged cells, where it was found to be mainly on the anode surface,[40,41,47,73,90] which is in accordance with simulations where the condition for Li deposition was first fulfilled also on the anode surface.[62,91,92]

The results of the NDP measurements allow more in-depth interpretations with regard to the positioning of the electrode sheets within the cell. The difference between Sheet 2-anode and Sheet 10-anode (*Fig. 14*) is caused by the different positions of the sheets within the electrode stack in the battery cell. Sheet 10 was extracted from the middle of the electrode stack, where, because of the higher temperature, less Li plating occurs on the anode surface. This behavior is similar to that of 26650 cells, where less Li plating was also found in the inner part of the jellyroll by GD-OES depth profiling.[41]

The middle position in both anodes, sheets 2 and 10, reveals an abrupt decrease in the Li concentration after the first μm. Right now, it is unclear if this finding is an artificial effect or a real depletion of Li. In summary, both studied Sheet-anodes display enrichment of Li at the sample surface assigned to Li plating in various degrees depending on one hand of the anode layer position in the electrode stack and on the other hand on the position (corner, edge, and middle) measured on a particular layer. The FIB image data, in combination with the EDX results, confirm the findings of the NDP measurements that there is a Li layer on the electrode. The NDP measurements thus provide valid information about the Li distribution in the depth of the material.

A preference for Li deposition on the anode surface was also recently shown by in situ optical microscopy of cross-sectioned full cells.[73]

Now that the detection of Li plating by the various methods has been provided and confirmed, and it is certain that the Li plating has only taken place on the surface of the sample, the results of the purely optical methods (scanner, microscopy) for the spatial resolution of the Li plating on the surface are considered below.

The imaging methods show that Li plating occurs next to the sheet edges at first (*Fig. 3*). These results confirm the results from.[10,58,93–95] In addition to that, most of the plating is next to the sheet edges; a preferred plating edge can be observed across all sheets - the side of the cathode tab. This is confirmed by both the scanner images and the light microscope images (*Fig. 4*). The light microscope images show a higher resolution and provide additional structural information about the deposition rounds so that needles can be recognized and their length measured. Deeper structural analyses are possible with the laser scanning microscope (*Fig. 6*). Here, additional depth information is collected in addition to the light image. This way makes it possible to recognize the thickness of the deposits, view 3D morphologies, and increase the magnification. The cost of the measurement devices increases. All three devices have the advantage of being able to be operated directly in the glovebox in an Ar atmosphere and produce results quickly without the need for sample preparation.

The Li plating distribution can be explained by inhomogeneous temperature distributions during the fast charging event. This assumption is supported by a significantly higher amount of Li plating on the outer electrodes compared to the ones in the middle of the stack. Since no cooling body was used, a higher cell core temperature is expected than for the outer surfaces, which are cooled via natural convection. Additionally, since the current of the entire anode electrode is collected through the anode tab (top right corner in the images), it is assumed that the anode tab side is slightly warmer compared to the cathode side. In general, inhomogeneous pressure distribution increases the plating risk.[96] Due to the pouch cell housing and the absence of mechanical clamping in this case, it can be assumed that the change in thickness of the electrodes does not lead to any significant change in pressure.

The literature shows that it is also possible to observe the anode during the Li plating process by setting up an optical cell.[73,97] These results show an optical cell that selectively deposits metallic Li. The deposition is not homogeneous in this cell either.

In order to transfer the results to the application, it is necessary to identify an externally measurable variable. Previous methods have shown that the deposits accumulate on the graphite surface and could be identified as metallic Li.

Therefore, optical cells, as in[73,97] were, for example, used.

In the following, the electrical signal of the cell before the cell opening is analyzed.

The cell analyzed here with Li plating on the anode shows the plateau in the DVA curve already known from the literature. The verification by post-mortem methods supports that the stripping peak is an indicator of the metallic Li deposition during the charging of the cell.

A broad investigation was carried out in,[59] with different C-rates and temperatures including post-mortem analysis and the image analysis also shown here, which also proves a qualitative correlation between the position of the minimum and the amount of platinum on the surface of the electrode.

**Conclusions**

We conclude that the methods used here allow for the investigation of Li plating. Depending on the initial situation and the ability to handle the battery, it is possible to detect Li-plating or to concentrate on the cell behaviour or material structures that can be assigned to Li-plating. All used methods are combined in *Fig. 15*. The employed EDX detector especially allows the detection of metallic Li. The EPR signal can distinguish between the different states of Li (intercalated or metallic).

NDP and GD-OES can resolve different amounts of Li at depth, whereas EDX has a spatial resolution. The presented work aims to provide a collection of methods for investigating metallic Li in batteries. It should be emphasized that all methods were carried out on the same material in the same cell, which made validating the methods against each other possible. The overview of the methods shows their strengths and weaknesses. With SEM-EDX, NDP, GD-OES, NMR, and EPR, Li can be detected spectroscopically. The optical techniques allow imaging over the area and depth of the sample. The optical methods differ in resolution and the associated area to be observed. The advantage of the optical methods is the possibility of observing the structure and, depending on the method, the color.

Many methods require additional evidence for verification to ensure that the amount of Li found is in metallic form. The optical methods show weaknesses in that they can only show the structures, but the evidence for the material is missing. The methods with a colored image have the additional indicator of the silver color for metallic Li. However, these two types of methods (optical and spectroscopic) can be combined and thus support the statement that Li plating is involved.

This work should serve as a basis for future publications to select a suitable method for their investigations and to be able to fall back on a validated method.

For future investigations, it is recommended to link an optical method with a spectroscopic method and to conduct direct Li detection in advance. This initial detection serves as a reference for subsequent measurements, ensuring that the focus of the actual investigation is not lost. With this work, many methods for Li detection were linked and made available as a reference for future laboratory work.

Importantly, all the findings from our research can now be leveraged to simulate Li plating. These simulations necessitate spatial data for validation and comprehension, as well as material composition data to ascertain the correct chemical reactions.

**Acknowledgments**

The authors gratefully acknowledge the funding of this work by the Federal Ministry of Education and Research (BMBF) of Germany in the project InOPlaBat (funding code: 03XP0352) and CharLiSiko (03XP0333). It is acknowledged that some of the key NDP experiments were carried out at the NPI and RC infrastructures "CANAM" and "Reactors LVR-15 and LR-0" in Řež with the support of the Ministry of Education, Youth and Sport of the Czech Republic (Projects No. LM2015056 and LM2015074). This work was also supported by the Ministry of Education, Youth and Sports (MEYS) CR under the project CZ.02.01.01/00/22_008/0004591.## References

[1.] Waldmann T., Hogg B.-I. and Wohlfahrt-Mehrens M. 2018 J. Power Sources **384** 107

[2.] Weiss M. et al. 2021 Adv. Energy Mater. **11**

[3.] Kondou H., Kim J. and Watanabe H. 2017 Electrochemistry **85** 647

[4.] Stottmeister D. and Groß A. 2023 Batteries & Supercaps **6**

[5.] Fleischhammer M., Waldmann T., Bisle G., Hogg B.-I. and Wohlfahrt-Mehrens M. 2015 J. Power Sources **274** 432

[6.] Waldmann T. and Wohlfahrt-Mehrens M. 2017 Electrochim. Acta **230** 454

[7.] Waldmann T., Quinn J. B., Richter K., Kasper M., Tost A., Klein A. and Wohlfahrt-Mehrens M. 2017 J. Electrochem. Soc. **164** A3154

[8.] Feinauer M., Abd-El-Latif A. A., Sichler P., Aracil Regalado A., WohlfahrMehrens M. and Waldmann T. 2023 J. Power Sources **570** 233046

[9.] Smart M. C. and Ratnakumar B. V. 2011 J. Electrochem. Soc. **158** A379

[10.] Ringbeck F., Rahe C., Fuchs G. and Sauer D. U. 2020 J. Electrochem. Soc. **167** 90536

[11.] Uhlmann C., Illig J., Ender M., Schuster R. and Ivers-Tiffée E. 2015 J. Power Sources **279** 428

[12.] von Lüders C., Zinth V., Erhard S. V., Osswald P. J., Hofmann M., Gilles R. and Jossen A. 2017 J. Power Sources **342** 17

[13.] Zinth V., von Lüders C., Hofmann M., Hattendorff J., Buchberger I., Erhard S., Rebelo-Kornmeier J., Jossen A. and Gilles R. 2014 J. Power Sources **271** 152

[14.] Ecker M., Shafiei Sabet P. and Sauer D. U. 2017 Appl. Energy **206** 934

[15.] Kühnle H., Knobbe E. and Figgemeier E. 2022 J. Electrochem. Soc. **169** 40528

[16.] Kayser S. A., Mester A., Mertens A., Jakes P., Eichel R.-A. and Granwehr J. 2018 Physical Chemistry Chemical Physics: PCCP **20** 13765

[17.] Niemöller A., Jakes P., Eichel R.-A. and Granwehr J. 2018 Sci. Rep. **8** 14331

[18.] Petzl M. and Danzer M. A. 2014 J. Power Sources **254** 80

[19.] Diaz L. A., Hnát J., Heredia N., Bruno M. M., Viva F. A., Paidar M., Corti H. R., Bouzek K. and Abuin G. C. 2016 J. Power Sources **312** 128

[20.] Jia H. and Xu W. 2022 Trends in Chemistry **4** 627

[21.] Thompson A. C. et al. 2001 X-ray Data Booklet: Center for X-ray Optics Advanced Light Source ()

[22.] Österreicher J. A., Simson C., Großalber A., Frank S. and Gneiger S. 2021 Scr. Mater. **194** 113664

[23.] Pecher O., Carretero-González J., Griffith K. J. and Grey C. P. 2017 Chem. Mater. **29** 213

[24.] Pell A. J., Pintacuda G. and Grey C. P. 2019 Prog. Nucl. Magn. Reson. Spectrosc. **111** 1

[25.] Schleker P. P. M., Grosu C., Paulus M., Jakes P., Schlögl R., Eichel R.-A., Scheurer C. and Granwehr J. 2023 Commun. Chem. **6** 113

[26.] Grey C. P. and Dupré N. 2004 Chem. Rev. **104** 4493

[27.] Blanc F., Leskes M. and Grey C. P. 2013 Acc. Chem. Res. **46** 1952

[28.] Leanza D., Vaz C. A. F., Czekaj I., Novák P. and El Kazzi M. 2018 J. Mater. Chem. A **6** 3534